\documentclass[superscriptaddress,aps,twocolumn,pre,10pt]{revtex4}
\usepackage[pdftex]{graphicx}

\usepackage{amsmath,amssymb}
\usepackage[latin1]{inputenc}
\usepackage[english]{babel}
\usepackage{xcolor}
\usepackage[pdftex]{hyperref}

\renewcommand{\epsilon}{\varepsilon}
\providecommand{\url}[1]{\texttt{#1}}

\begin{document}
\author{M. E. Cates}
\author{D. Marenduzzo}
\affiliation{SUPA, School of Physics and Astronomy, University of
Edinburgh, Mayfield Road, Edinburgh EH9 3JZ, UK}
\author{I. Pagonabarraga}
\affiliation{Departament de F\'{\i}sica Fonamental, Universitat de Barcelona - Carrer Mart\'{\i} Franqu\'es 1, 08028-Barcelona, Spain}, 
\author{J. Tailleur}
\affiliation{SUPA, School of Physics and Astronomy, University of
Edinburgh, Mayfield Road, Edinburgh EH9 3JZ, UK}

\begin{abstract} 
  We present a generic mechanism by which reproducing microorganisms,
  with a diffusivity that depends on the local population density, can
  form stable patterns. It is known that a decrease of swimming speed
  with density can promote separation into bulk phases of two
  coexisting densities; this is opposed by the logistic law for birth
  and death which allows only a single uniform density to be
  stable. The result of this contest is an arrested nonequilibrium
  phase separation in which dense droplets or rings become separated
  by less dense regions, with a characteristic steady-state length
  scale. Cell division mainly occurs in the dilute regions and cell
  death in the dense ones, with a continuous flux between these
  sustained by the diffusivity gradient. We formulate a mathematical
  model of this in a case involving run-and-tumble bacteria, and make
  connections with a wider class of mechanisms for density-dependent
  motility. No chemotaxis is assumed in the model, yet it predicts the
  formation of patterns strikingly similar to those believed to result
  from chemotactic behavior.
\end{abstract}
\date{\today}
 
\title{Arrested phase separation in reproducing bacteria: a generic route to pattern formation?}

\maketitle

Microbial and cellular colonies are among the simplest examples of
self-assembly in living organisms. In nature, bacteria are often found
in concentrated biofilms, mats or other colony types, which can grow
into spectacular patterns visible under the
microscope~\cite{shapiro,harshey}. Also in the laboratory, bacteria
such as {\em E. coli} and {\em S. typhimirium} form regular geometric
patterns when they reproduce and grow on a Petri dish containing a gel
such as agar.  These patterns range from simple concentric rings to
elaborate ordered or amorphous arrangements of
dots~\cite{budrene1,budrene2,woodward,murray}.  {Their formation
  results from collective behaviour driven by interactions between the
  bacteria, such as chemotactic aggregation~\cite{murray}, competition
  for food~\cite{kawasaki} or changes in phenotypes according to
  density~\cite{espiov}.} The question as to whether general
mechanisms {lie behind} this diversity of {microscopic pathways to
  patterning} remains open.

Unlike the self-assembly of colloidal particles, pattern formation in
motile microorganisms and other living matter is typically driven by
non-equilibrium rather than thermodynamic forces. Indeed, the dynamics
of both dilute and concentrated bacterial fluids is already known to
be vastly different from that of a suspensions of Brownian
particles. For instance, suspensions of active, self-propelled
particles, have been predicted to exhibit giant density
fluctuations~\cite{toner,toner2}, which have been observed
experimentally~\cite{sriram}. Similarly, an initially uniform
suspension of self-propelled particles performing a ``run-and-tumble''
motion like {\it E. coli} has recently been shown theoretically to
separate into a bacteria-rich and a bacteria-poor phase, provided that
the swimming speed decreases sufficiently rapidly with
density~\cite{tailleur}.  This is akin to what happens in the spinodal
decomposition of binary immiscible fluids, but has no counterpart in a
system of Brownian particles interacting solely by density-dependent
diffusivity. (The latter obey the fluctuation-dissipation theorem,
ensuring that the equilibrium state is diffusivity-independent.) Other
non-equilibrium effects, such as ratchet physics, have also been
observed and used either to rectify the density of
bacteria~\cite{Galajda,Galajda2,Tailleurepl} or to extract work from
bacterial assemblies~\cite{Ruocco}.

Some aspects of bacterial patterning show features common to other
nonequilibrium systems, and a crucial task is to identify the key
ingredients that control their development. In many equilibrium and
nonequilibrium phase transitions an initial instability creates
density inhomogeneities; these coarsen, leading eventually to
macroscopic phase separation \cite{chaikin}. The situation observed in
bacterial assemblies often differs from this; long-lived patterns
emerge with fixed characteristic length scales, suggesting that any
underlying phase separation is somehow arrested.  The strong diversity
of biological functions met in experiments has led to an equally
diverse range of proposed phenomenological
models~\cite{woodward,murray,benjacob,tyson,brenner,espiov} to account
for such effects.  Most of them rely on the coupling of bacteria with
external fields (food, chemoattractant, stimulant, etc.), and many
involve a large number of parameters due to the complexity of the
specific situation of interest. The most common mechanism used to
explain the bacterial patterns is chemotaxis~\cite{murray}: the
propensity of bacteria to swim up/down gradients of
chemoattractants/repellants. This explanation is so well established
in the literature for at least two organisms ({\em E. coli} and {\em
  S. typhimuridium} \cite{murray}) that observation of similar
patterns in other species might defensibly be taken as evidence for a
chemotactic phenotype.

Here we identify a very general mechanism that can lead to pattern
formation in bacterial colonies and which may encompass a large class
of experimental situations. This mechanism involves a
density-dependent motility, giving rise to a phase separation which is
then arrested, on a well defined characteristic length scale, by the
birth and death dynamics of bacteria. For definiteness we will work
this through for a particular model of bacterial run-and-tumble
motion, involving a swim speed that depends (via unspecified
interactions) on local bacterial density. This gives pattern similar
to those observed in experiments~\cite{woodward,murray}. However, the
basic mechanism ---density-dependent motility coupled to logistic
population growth--- is not limited to this example. Our work
demonstrates that chemotaxis {\em per se} is not a prerequisite for
observing what are sometimes colloquially referred to as `chemotactic
patterns'.

It is indeed remarkable that density-dependent swim speed and logistic
growth alone are sufficient to create some of the specific pattern
types previously identified with specific chemotaxis mechanisms.  In
mechanistic terms, we find that the logistic growth dynamics
effectively arrests a spinodal phase separation that is known to
follow from a density-dependent swim speed ~\cite{tailleur}. Put
differently, an initially uniform bacterial population with small
fluctuations will aggregate into droplets, but these will not coarsen
further once a characteristic length scale is achieved, at which
aggregation and birth/death effects come into balance. Starting
instead from a small inoculum, we predict formation of concentric
rings which, under some conditions, at least partially break up into
spots at late times \cite{woodward}.

To exemplify our generic mechanism we will start from a minimal model
of run-and-tumble bacteria, that can run in straight lines with a swim
speed $v$ and randomly change direction at a constant tumbling rate
$\tau^{-1}$~\cite{schnitzer,berg}. To this we add our two key
ingredients: a local density-dependent motility, and the birth/death
of bacteria, the latter accounted for through a logistic growth
model. Of course, bacteria can interact locally in various ways,
ranging from steric collisions ~\cite{tailleur} to chemical
quorum-sensing~\cite{murray}. (Indeed a nonspecific dependence of
motility on bacterial density was previously argued to be central to
bacterial patterning by Kawasaki~\cite{kawasaki}.)  Here we focus on
the net effect of all such interactions on the swim speed $v(\rho)$,
which we assume to decrease with density $\rho$. This dependence might
include the local effect of a secreted chemoattractant (such as
aspartate \cite{budrene1,budrene2,woodward} which causes aggregation,
effectively decreasing $v$) but does not assume one.

In addition to their run-and-tumble motion, real bacteria continuously
reproduce, at a medium-dependent growth rate which ranges from about
one reciprocal hour in favourable environments such as {\it Luria
  broth} to several orders of magnitude lower for `minimal' media such
as M9. In bacterial colonies patterns may evolve on timescales of
days~\cite{murray}, over which such population growth dynamics can be
important.

We now derive continuum equations for the local density $\rho({\bf
  r},t)$ in a population of run-and-tumble bacteria, with swim speed
$v(\rho)$, growing at a rate $\alpha (1-\rho/\rho_0)$. The latter
represents a sum of birth and death terms, in balance only at $\rho =
\rho_0$.  At large scales in a uniform system, the motion of
individual bacteria is characterized by a diffusivity
$D(\rho)=v(\rho)^2 \tau/d$, where $\tau^{-1}$ is the tumbling rate and
$d$ the dimensionality~\cite{schnitzer,berg}. Crucially however, a
non-uniform swimming speed $v({\bf r})$ also results in a mean drift
velocity $V = -v\tau\nabla v$~\cite{schnitzer} which here gives $V
=-D'(\rho) \nabla \rho/2$ \cite{tailleur}.  This contribution is
crucial to phase separation~\cite{tailleur} and will again play a
major role here. However this term has no counterpart in ordinary
Brownian motion (even if particles have variable diffusivity) and was
accordingly overlooked in previous studies which relied on
phenomenological equations involving a density-dependent diffusivity
and no drift~\cite{kawasaki}.

Coupling the diffusion-drift equation for run-and-tumble bacteria, as
derived in~\cite{tailleur}, with the logistic growth term, the full
dynamics is then given by:
\begin{eqnarray}\label{1Ddynamics}
  \frac{\partial \rho({\bf r},t)}{\partial t} & = & \nabla\cdot 
  \left[ {\cal D}_{\rm e}(\rho) \nabla\rho({\bf r},t)\right]\\ \nonumber
  &  &+ \alpha \rho({\bf r},t)\left(1-\frac{\rho({\bf r},t)}{\rho_0}\right) 
  -\kappa  \nabla^4 \rho({\bf r},t)
\end{eqnarray}
where the `effective diffusivity' is 
\begin{equation}
  \label{eqn:Deff}
  {\cal D}_{\rm e}(\rho)=D(\rho)+\frac 1 2 \rho D'(\rho)
\end{equation}
This results from the summed effects of the true diffusive flux
$-D(\rho)\nabla \rho$ and the non-linear drift flux $\rho V$. In
Eq.\eqref{1Ddynamics} we have also introduced a phenomenological
surface tension $\kappa>0$, which controls gradients in the bacterial
density. Such a contribution has been shown to arise when the speed of
a bacterium depends on the average density in a small local region
around it, rather than a strictly infinitesimal one~\cite{tailleur}.
Eq.~\eqref{1Ddynamics} neglects noise, both in the run-and-tumble
dynamics and in the birth/death process. The former noise source
conserves density and should become irrelevant at the experimental
time scale of days. On the other hand, the non-conservative noise in
the birth and death dynamics may be more important, and we have
verified that our results are robust to its introduction at small to
moderate levels. Numerical simulations of Eq.~\eqref{1Ddynamics} have
been performed with standard finite difference methods (although noise
does require careful treatment, as in \cite{noise1,noise2}), with periodic
boundary conditions used throughout.  For definiteness, all our
simulations have been carried out with $v(\rho)=v_0 e^{-\lambda
  \rho/2}$, where $v_0>0$ is the swim speed of an isolated bacterium
and $\lambda>0$ controls the decay of velocity with density. The
precise form of $v(\rho)$ is however not crucial for the phenomenology
presented here, and the instability analysis offered below does not
assume it.

The logistic population dynamics alone would cause the bacterial
density to evolve toward a uniform density, $\rho({\bf r})=\rho_0$,
which constitutes a fixed point for the proposed model. Although this
homogeneous configuration is stable in the absence of bacterial
interactions, it has been shown~\cite{tailleur} that, without logistic
growth, a density-dependent swim speed $v(\rho)$ leads to phase
separation via a spinodal instability whenever $dv/d\rho < -v/\rho$.
By Eq.~\eqref{eqn:Deff} this equates to the condition ${\cal D}_{\rm
  e} < 0$, and it is indeed obvious that the diffusive part of
Eq.~\eqref{1Ddynamics} is unstable for negative ${\cal D}_{\rm e}$.
It is important, clearly, that ${\cal D}_{\rm e}$ can be negative
although $D$ is not. This holds for a much wider class of
nonequilibrium models than the one studied here; we return to this
point at the end of the paper.

For the choice of $v(\rho)$ made in our simulations, we have ${\cal
  D}_{\rm e} = D(\rho)[1-\rho\lambda/2]$ and the flat profile will
thus become unstable for $\rho_0$ above $2/\lambda$. We have confirmed
this numerically, and find that upon increasing $\rho_0$, the uniform
state becomes (linearly) unstable, evolving in a 1D geometry into a
series of ``bands'' of high bacterial density separated by low density
regions. Depending on the parameters, this transition can be
continuous (supercritical), with the onset of a harmonic profile whose
amplitude grows smoothly with $\rho_0$, or discontinuous (subcritical)
with strongly anharmonic profiles (see Fig. 1).

\begin{figure}[ht]
\hspace{-.65cm}  \includegraphics[width=\columnwidth]{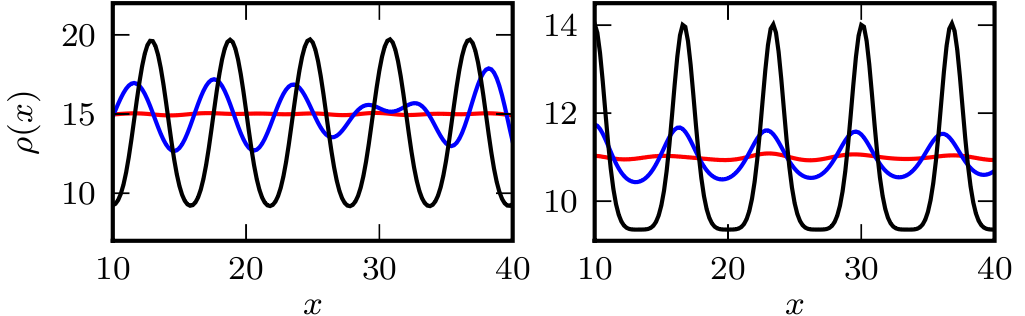}
  \caption{Growth of the instability in the supercritical (left) and
subcritical cases (right). The three lines correspond to three
successive times. A small perturbation around $\rho_0$ (red line)
growth toward harmonic or anharmonic patterns in the supercritical or
subcritical case, respectively. {\bf Left}: Supercritical case
($\alpha=\kappa=0.01$, $\lambda=0.02$, $\rho_0=15$, $D_0=1$, times:
$10^2$, $10^3$, $10^4$). {\bf Right}: Subcritical case
($\alpha=\kappa=0.005$, $\lambda=0.02$, $\rho_0=11$, $D_0=1$,
times: $3.\,10^2$, $3.\,10^3$, $10^5$)}
%\label{fig:disprelat}
\end{figure}

The transition to pattern formation arising from
Eq.~\eqref{1Ddynamics} is a fully nonequilibrium one: it is not
possible to write down an effective thermodynamic free energy which
would lead to this equation of motion.  Nonetheless, it is possible to
understand why the birth/death process effectively arrests the
spinodal decomposition induced by the density-dependent swim
speed. The latter tends to separate the system into high and low
density domains with densities either side of $\rho_0$. (Without the
logistic term, these would coarsen with time.) Bacteria thus tend to
be born in the low density regions and to die in the high density
regions. To maintain a steady state, they have to travel from one to
the other: balancing the birth/death terms by the diffusion-drift
transport flux between the domains then sets a typical scale beyond
which domain coarsening can no longer progress.  Were any domain to
become much larger, the density at its centre would soon regress
towards $\rho_0$, re-triggering the spinodal instability
locally. (This is closely reminiscent of what happens in a
thermodynamic phase separation when the supersaturation is
continuously ramped \cite{philtrans}.)

\begin{figure}[ht]
  \includegraphics[width=.8\columnwidth]{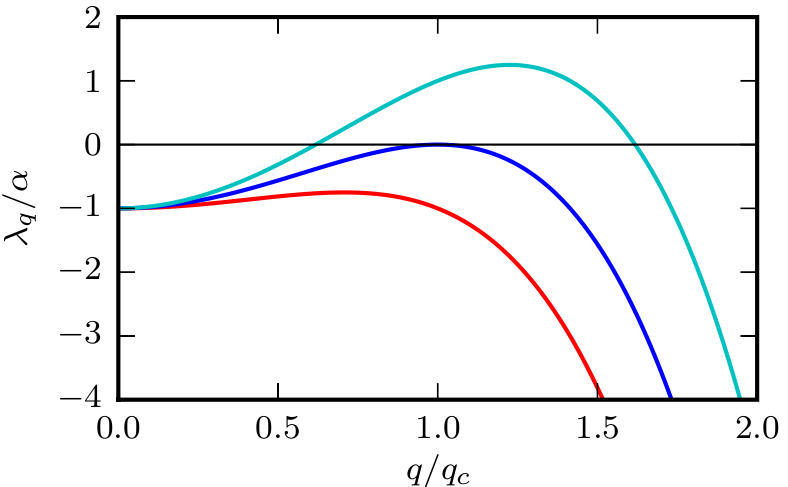}
  \caption{Three plots of $\Lambda_q(q)$ for ${|
    {\cal D}_{\rm e}(\rho_0)|}/{\sqrt{\alpha \kappa}}=1;2;3$. At the transition, only
    one critical mode $q=q_c$ is unstable.}
%  \label{fig:disprelat}
\end{figure}

To better understand the onset of the instability, let us linearize
Eq.~\eqref{1Ddynamics} around $\rho({\bf r})=\rho_0$ and work in
Fourier space. Defining $\rho({\bf r})=\rho_0+\sum_q \delta \rho_q
\exp(i{\bf q}\cdot{\bf r})$ yields:
\begin{equation}
  \label{eqn:Lambdaq}
  \dot \delta \rho_q =  \Lambda_q \delta_q;\qquad \Lambda_q= -\alpha-q^2 {\cal D}_{\rm e}(\rho_0)-\kappa q^4
\end{equation}
The flat profile $\rho=\rho_0$ is thus stable if $\Lambda_q \leq 0$
for all $q$ and is unstable otherwise. From the expression for ${\cal
  D}_{\rm e}(\rho_0)$, Eq.~\eqref{eqn:Deff}, one sees that instability
occurs if
\begin{equation}
\label{eqn:critunst}
\Phi\equiv-\frac{\rho_0 D'(\rho_0)}{2D(\rho_0)}\geq 1\quad\mbox{and}\quad -\frac{{\cal D}_{\rm e}(\rho_0)}{\sqrt{\alpha \kappa}}\geq 2
\end{equation}
At the onset of the instability only one mode is unstable, with
wavevector $q_c=\sqrt{{2\alpha}/{|{\cal D}_{\rm e}(\rho_0)|}}$, as can
be seen in Figure 2.  The first condition in Eq.~\eqref{eqn:critunst},
$\Phi\geq 1$, is equivalent to the requirement that ${\cal D}_{\rm
  e}<0$ given previously. From the dispersion relation,
Eq.~\eqref{eqn:Lambdaq}, we see that the resulting destabilization is
balanced by the stabilizing actions of bacterial reproduction and the
surface tension at large and small wavelength, respectively. The
unstable modes thus lie within a band $q_1 < q < q_2$ where $q_1\simeq
q_\alpha\equiv\sqrt{\alpha/| {\cal D}_{\rm e}(\rho_0)|}$ and
$q_2\simeq q_\kappa\equiv\sqrt{| {\cal D}_{\rm e}(\rho_0)|/\kappa}$
set the wavelengths below and above which the stabilizing effects of
bacteria reproduction and the surface tension can compete with the
destabilizing effect of the negative diffusivity, respectively. For
unstable modes to exist, one needs $q_1\le q_2$; restoring prefactors,
this yields $2q_\alpha\leq q_\kappa$ which is the second criterion in
Eq.~\eqref{eqn:critunst}.  This analysis is consistent with the view
that phase separation is arrested by the birth/death dynamics, which
stabilizes the long wavelength modes ($\Lambda_0=-\alpha$), while the
phenomenological tension parameter $\kappa$ primarily fixes the
interfacial structure of the domains, not their separation.

\begin{figure}[ht]
  \begin{center}
    \includegraphics{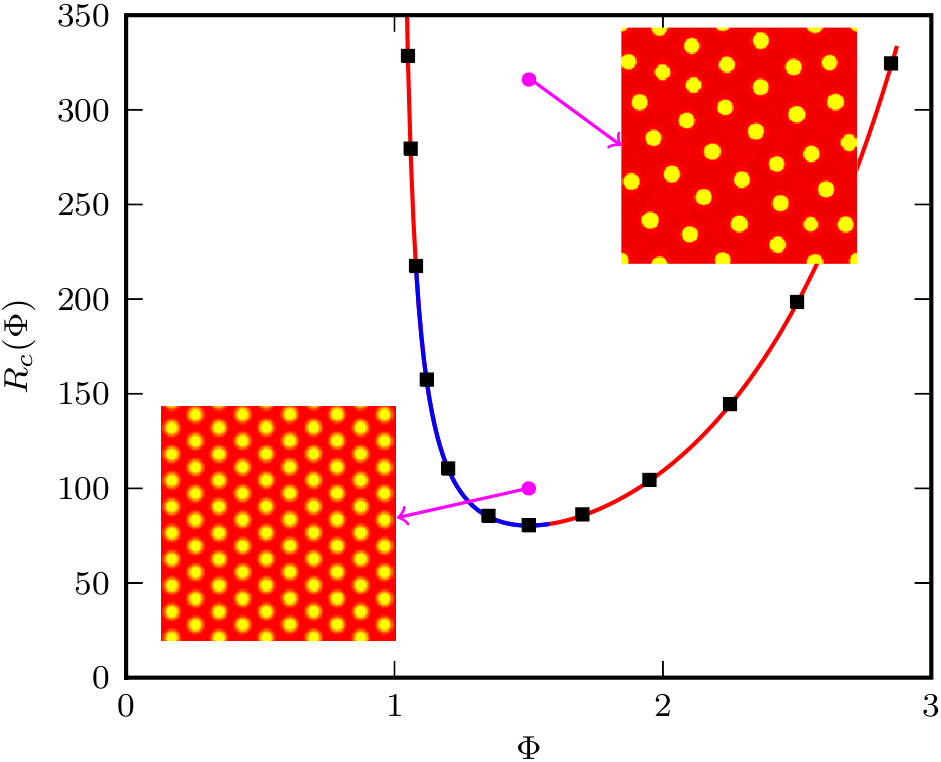}\\
    \includegraphics{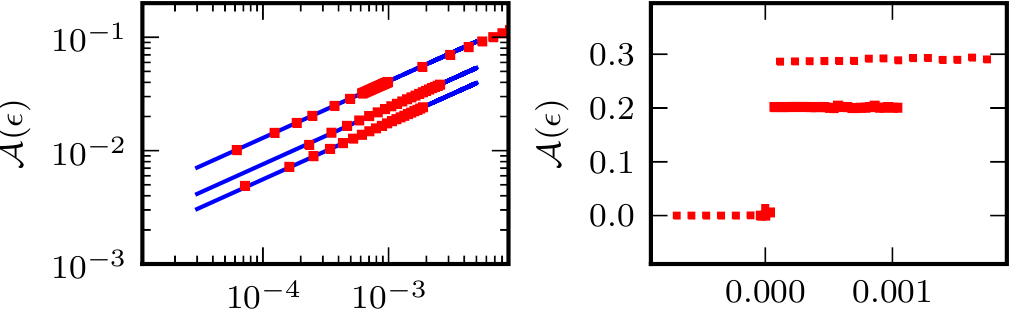}
  \end{center}
  \caption{{\bf Top}: Phase diagram in the $(R,\Phi)$ plane. The outer
    region corresponds to stable behavior whereas within the curve,
    patterning occurs. The solid line is the theoretical phase
    boundary -- Eq \eqref{eqn:PD} -- which accurately fits the
    numerics (black squares). The blue and red sections correspond to
    continuous and discontinuous transitions respectively. The two
    magenta dots correspond to two 2D simulations which show ordered
    harmonic patterns close to supercriticality and amorphous patterns
    otherwise.  {\bf Bottom-left}: Transition in the supercritical
    regime. The blues lines correspond to the theory -- Eq
    \eqref{eqn:amp} -- whereas the squares comes from simulations
    ($\Phi=1.5;\ 1.35;\ 1.2$ from top to bottom).  {\bf Bottom-right}:
    Transition in the subcritical regime for $\Phi=1.06$ and
    $\Phi=1.7$ (bottom to top).}
  \label{fig:PD}
\end{figure}

We now consider more closely the parameters controlling the transition
to pattern formation. For definiteness, we address the specific case
used for our simulations, $D(\rho)=D(0)\exp(-\lambda \rho)$. To put
Eq.~\eqref{1Ddynamics} in dimensionless form, we define rescaled time,
space and density as
\begin{equation}
  \tilde t=\alpha t;\;\; \tilde{\bf r}= \left(\frac\alpha\kappa\right)^{1/4}
\!\!  {\bf r};\;\; u=\frac\rho{\rho_0}
\end{equation}
 The equation of motion now reads
\begin{equation}
  \label{eqn:rescaled}
  \dot u= \nabla \cdot [ R e^{-2 \Phi u}(1-\Phi u)\nabla u]+u(1-u)-\nabla^4 u
\end{equation}
where $R\equiv D_0/\sqrt{\alpha\kappa}$ and $\Phi= \lambda\rho_0/2$
are the two remaining dimensionless control parameters.  Meanwhile the
conditions \eqref{eqn:critunst} for pattern formation become
\begin{equation}
  \label{eqn:PD}
  \Phi\ge 1;\qquad\qquad R \geq R_c= 2 \frac{\exp(2\Phi)}{\Phi-1}
\end{equation}
These relations, combined with the preceding linear stability
analysis, define a phase diagram in the $(R,\Phi)$ plane (Figure
\ref{fig:PD}) that agrees remarkably well with numerical results for
systems prepared in a (slightly noisy) uniform initial state.

Close to the transition, the emergent steady-state pattern can be
studied using an amplitude equation (see Supporting Information).
Introducing $\epsilon= (R-R_c)/R_c$, one gets in 1D that for $1.08\leq
\Phi \leq 1.58$, the transition is supercritical {(continuous)} and
the steady state is given by
\begin{equation}
\label{eqn:amp}
  \begin{aligned}
    u&\simeq 1+{\cal A}(\epsilon) \cos(x);\\
    {\cal A}^2(\epsilon)&=\epsilon\,\frac{18 (1-\Phi)^2}{34\Phi^4-56 \Phi^3-24 \Phi^2+31 \Phi+19}
  \end{aligned}
\end{equation}
which agrees with simulations (Fig 3, bottom left). Outside this
range, the transition becomes subcritical ({discontinuous,} Fig 3,
bottom right) and the analytical tools available become less
reliable~\cite{alex}. We emphasize again that the basic mechanism for
patterning presented above does not depend on the precise form chosen
for $v(\rho)$. Quantitatively however, Eq.~\eqref{eqn:amp}, and the
frontier between subcriticality and supercriticality, do depend on the
details of the interplay between the nonlinearity in $v(\rho)$ and the
logistic growth term. We leave further analysis of such model-specific
features to future work.

\begin{figure*}
\centering {\includegraphics[width=12cm]{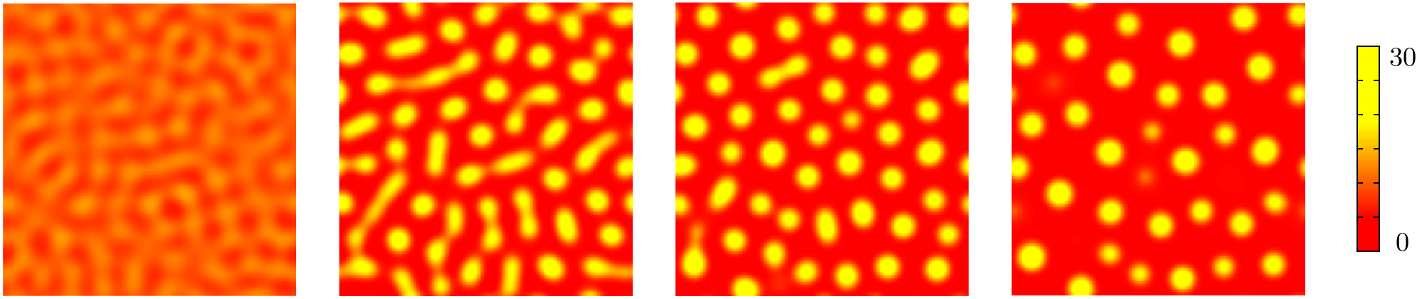}}
\caption{Numerical results for a 2D simulation with
size equal to 16$\times$16, $\lambda=0.3$, $D_0=1$, 
$\alpha=0.01$, $\kappa=0.001$ and $\rho_0=10$. 
Times corresponding to the snapshots are (in simulation units, from
left to right): 5, 12, 19.6 and 50.6.}
\end{figure*}

While the amplitude equation is more easily developed in 1D, the
stability analysis offered above is valid in higher dimensions and it
is natural to ask what happens in 2D, which is the relevant geometry
for Petri dish studies with growing bacterial colonies. Fig. 4 shows
the simulated time evolution of $\rho({\bf r},t)$ for a system started
with small random fluctuations around the equilibrium density
$\rho_0$, with other parameters as in Fig. 1. Perhaps not
surprisingly, bands are replaced by droplets of the high density phase
dispersed in a low density background at large times. This is the
typical steady state obtained with a near-uniform starting
condition. However, the structure and organization of the bacterial
drops in the steady state depends on the point ($R,\Phi$) chosen in
the phase diagram. Generally, the closer the system is to the
supercritical instability curve, the more ordered the patterns. For
instance we have observed an essentially crystalline distribution of
bacterial drops, which develops defects and eventually becomes
{amorphous} on moving further away from the phase boundary (Fig. 3,
insets to main panel). For particular choices of parameters, our model
can also admit other steady state patterns. Close to the supercritical
line, where the phase transition is continuous, we can obtain
long-lived stripes, whereas for fixed large values of $R$ and $\Phi$
close to the (right) subcritical phase boundary, we have also observed
`inverted droplets' with a high density lawn punctuated by low density
`holes'.

In these 2D geometries initialized from a near-uniform state, droplets
can coalesce in the early stages, while at late times the dynamics is
governed by evaporation-condensation events (see Figs. 4 and 5).
However it is already apparent from Fig. 4 that coarsening eventually
stops and the droplets reach rather well-defined steady state sizes
and centre-to-centre distances. This can be quantified by looking at
the time evolution of the characteristic domain size, $L(t)$, which we
have computed as the inverse of first moment (times $2\pi$) of the
structure factor~\cite{chaikin}. Fig. 5 suggests that $L(t)$ at late
times eventually stops increasing and reaches a steady state
value. (The visible steps in domain size mark discrete evaporation
events involving smaller bacterial droplets; presumably $L(t)$ would
become smooth for a large enough system.)

 {\begin{figure}
    \includegraphics[width=7cm]{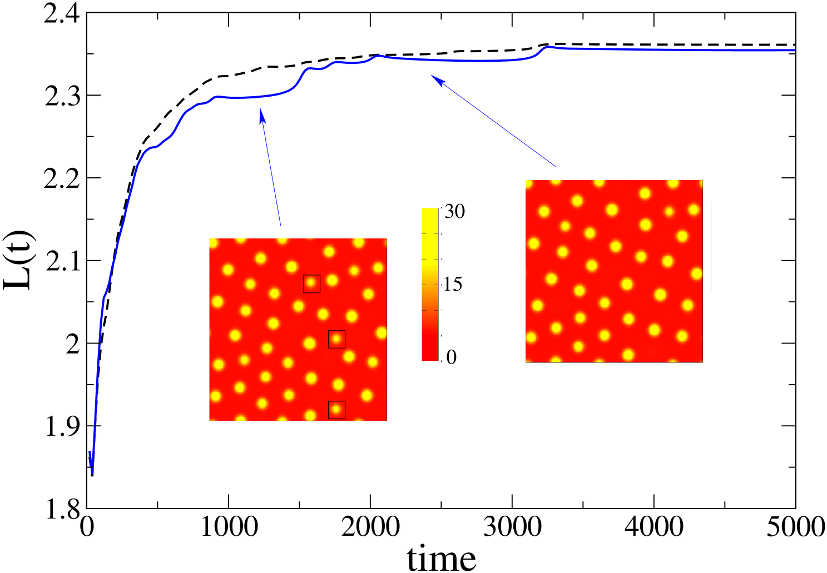}
  \caption{Plot of the characteristic domain size, $L(t)$ as a function
    of time for a system in the inhomogeneous phase, with initially
    random density fluctuations around $\rho_0$. Parameters were:
$\alpha=0.01$, $\lambda=0.27$, $\kappa=0.001$, while the system size
was $20\times 20$. The solid line corresponds to a single run, while
the dashed line is an average over 6 runs. The steps in the single 
run curve correspond to evaporation-condensation events, highlighted by
black squares in the snapshots shown in the figure (before and after one
of the steps respectively, arrows indicate positions on the plot corresponding
to the two snapshots).}
\end{figure}}

These droplet patterns in steady state are very similar to those
observed for {\em E. coli} in a liquid medium or {\em S. typhimurium}
in semi-solid agar (0.24\% water-agar in Ref.~\cite{woodward}) when
starting from a uniform distribution~\cite{murray}. For the {\em E.
  coli} case, interactions are believed to come from chemoattractant,
emitted by the bacteria themselves, that is not degraded over time
\cite{budrene1,budrene2}. The chemoattractant distribution should
become more and more uniform so that these interactions decay to zero
as time proceeds. In our framework this is analagous to decreasing
$\Phi$, which will turn any initially unstable state into a
homogeneous one, and can thus explain that the patterns observed
experimentally fade with time (whereas in our simulations $\Phi$
remains constant and the pattern are stable indefinitely). {\em
  E. coli} in a semi-solid medium also exhibits droplet patterns of
high symmetry. In our framework, such patterns result from a
continuous transition, close to the supercritical line.

The growth of bacterial colonies of {\it S. typhimurium} starting from
a small inoculum of bacterial cells in semi-solid agar\if{on a Petri
  dish}\fi leads to quite specific (transient but long-lived)
patterns, with the bacteria accumulating in concentric rings that can
subsequently fragment into a pattern of dots~\cite{woodward,murray}.
Once again, although these patterns are believed to stem from a
chemotactic mechanism ~\cite{murray}, we find they can arise in
principle without one, so long as our two basic ingredients of
density-suppressed motility and logistic growth are both present.
Indeed, initializing our simulations with a single small droplet of
high density $\rho$, we find that a similarly patterned bacterial
colony structure develops.  First, the bacteria spread radially
(through a Fisher-like wave), forming an unstructured lawn with the
highest density at the center.  This background density increases
logistically until the onset of instability via our generic
phase-separation mechanism; with circular symmetry, the instability
causes concentric rings of high bacterial density to successively
develop that are very stable in time (Figs. 6c and 6d).  The patterns
observed at later times again depend on the position of the parameters
in the $(R,\Phi)$ plane.  If we fix a value of $R$, e.g. 100, larger
values of $\Phi$ in the unstable region lead to rings being very
stable. For smaller values of $\Phi$, on the other hand, effectively
corresponding to weaker interactions between the bacteria, we observe
that rings initially form but rapidly destabilize through a secondary
modulation of the bacterial density along them.  This eventually
breaks the rings into a series of drops. The inner rings destabilize
first, and the system evolves eventually to the same steady state as
found starting from a uniform density, composed of drops with well
defined characteristic size and separation.  All this phenomenology is
strikingly reminiscent of the dynamics observed by Woodward {\em et
  al}~\cite{woodward} for {\it S. typhimirium}, where rings are stable
at large concentrations of potassium succinate (a `stimulant' which
promotes pattern formation), but break up into drops at smaller ones.
Our model shows a similar morphological change when decreasing $\Phi$,
i.e. the strength of the interactions.\if{ on going from the centre to
  the edge of the pattern-forming region on the phase diagram.}\fi

\begin{figure*}
\centering  \includegraphics[width=12cm]{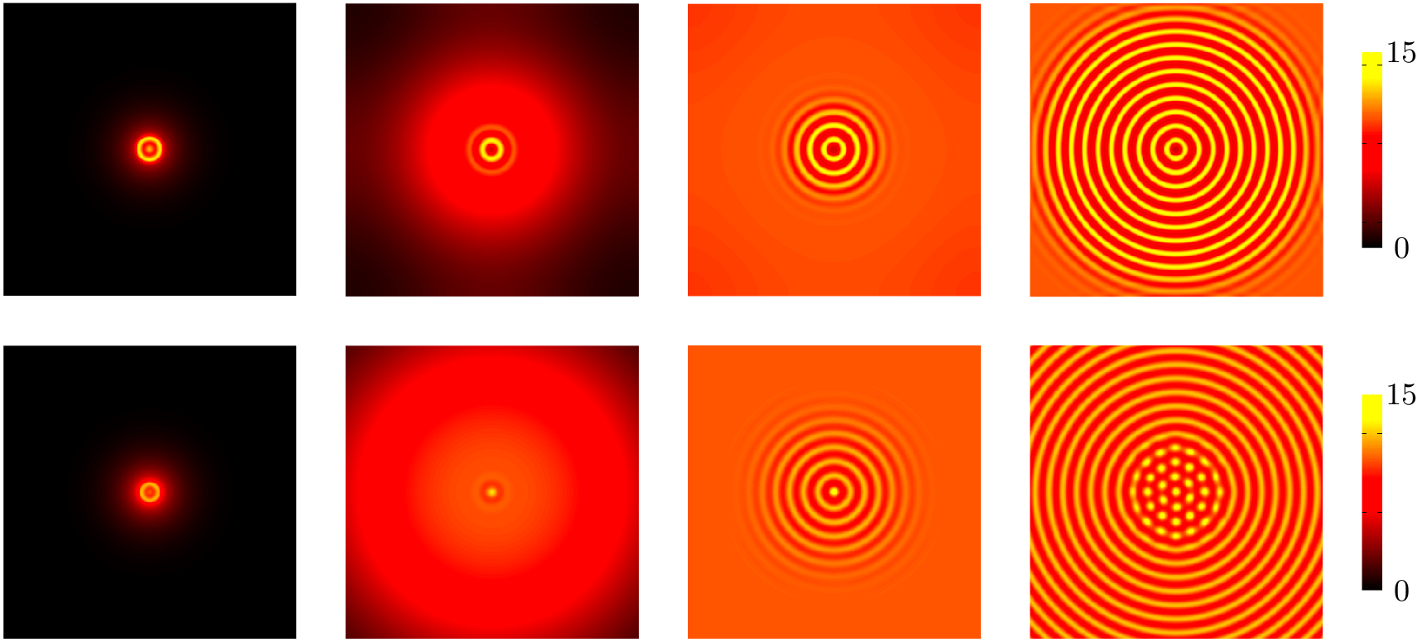}
  \caption{Dynamics of formation of "chemotactic patterns" in 2D,
    starting from a single small bacterial droplet in the middle of
    the simulation sample.  Top row: Formation of "chemotactic rings"
    in a system with $\alpha=0.1$, $\lambda=0.33$, and
    $\kappa=0.001$. The simulation box has size $40\times 40$. The
    snapshots correspond to times equal to (in simulation units, from
    left to right) 10, 50, 100 and 270.  Bottom row: Breakage of
    "rings" into "dots". The four snapshots correspond to the time
    evolution of a system with $\alpha=0.1$, $\lambda=0.26$, and
    $\kappa=0.001$. We show a $40\times 40$ fraction of the simulation
    box, with the boundaries far away and not affecting the pattern.
    The snapshots correspond to times equal to (in simulation units,
    from left to right) 10, 70, 290 and 1220.}
\end{figure*}

Different views are possible concerning the ability of our generic
model to reproduce the observed chemotactic patterns of {\em E. coli}
and {\em S. typhimurium} \cite{murray}. One possibility is that
Eq.~\eqref{1Ddynamics}, with the interpretation we have given for it,
actually does embody the important physics of pattern formation in
these organisms. Indeed it is well accepted that bacteria in the high
density concentric rings are essentially non-motile~\cite{mittal}. The precise
mechanism leading to this observation is not well
understood~\cite{murray}, but it is possible that the chemotactic
mechanism mainly acts to switch off motility at high density. If so,
by focussing solely on this aspect (with a correspondingly vast
reduction in the parameter space from that of explicit chemotactic
models \cite{woodward,murray}) our model might capture the physics of
these chemotactic patterns in a highly economical way. Interestingly,
our model is essentially local, whereas chemotaxis in principle
mediates interactions between bacteria that are nonlocal in both space
and time. It is not clear whether such nonlocality is essential for
the chemotactic models in~\cite{woodward,murray} or if fast-variables
approximations and gradient expansions would reduce these models
(which invole between 9 and 12 parameters) into Eq.[1]. In this case,
we would still have in Eq.[1] a highly economical model for
chemotactic pattern-formation organisms, possibly with a different
interpretation of ${\cal D}_{\rm e}$ and $\kappa$.

Alternatively, the success of our local model for these chemotactic
organisms might be largely coincidental. But in that case, such a
sparse model should be easily falsified, for instance by using the
linear stability analysis to relate the typical length scale of the
patterns to microbial parameters. This length scale is of order
$2\pi/q_c=2\pi \sqrt{|{\cal D}_{\rm e}|/\alpha}$, with $|{\cal D}_{\rm
  e}| \simeq D$, a typical bacterial diffusion coefficient
($D\sim{\cal O}$(100 $\mu$m$^2$s$^{-1}$) for {\it E.
  coli}~\cite{berg}). Using the previously quoted growth rate
$\alpha\simeq 1$ hr$^{-1}$, we get a ring separation of $\sim$ 1 mm,
in order-of-magnitude agreement with the experimental
value~\cite{woodward}.  This test could perhaps be sharpened usefully
by altering the growing medium so as to change $\alpha$.

More generally, our analysis of Eq.~\eqref{1Ddynamics} shows that the
main prerequisite for pattern formation, assuming the presence the
logistic growth term, is negativity of the effective diffusion
constant ${\cal D}_{\rm e}$. For run-and-tumble dynamics, ${\cal
  D}_{\rm e}<0$ was shown to arise for a sufficiently strong decay of
swim-speed with density; it does so because spatial variations in the
true diffusivity $D(\rho)$ create a drift flux $\rho V= -\rho
D'(\rho)\nabla\rho/2$ which can overcompensate the true diffusive flux
$-D\nabla\rho$~\cite{tailleur}. Negative ${\cal D}_{\rm e}$ could,
however, equally arise for any density dependent nonequilibrium
diffusion process. Indeed, the principle of detailed balance, which
holds only for equilibrium systems, leads to the Einstein relation,
that $D = k_BT\mu$ with $D$ a many-body diffusivity and $\mu$ the
corresponding mobility. This alone ensures that no drift velocity can
arise purely from gradients in $D$. In nonequilbrium systems, one
should expect generically to find such drift velocities, and the
run-and-tumble model is merely one instance of this. Accordingly one
can expect in principle to find cases of negative ${\cal D}_{\rm e}$
in other microorganisms showing distinctly different forms of
density-dependent self-propulsion.

To summarize, we have studied the dynamics of a system of reproducing
and interacting run-and-tumble bacteria, in the case where
interactions lead to a decreasing local swim speed with increasing
local density.  We have thereby identified a potentially generic
mechanism for pattern formation in which an instability towards phase
separation, caused by the tendency for bacteria to move slowly where
they are numerous, is arrested by the birth and death dynamics of
bacterial populations. We have shown that these two ingredients alone
are enough to capture many of the patterns observed experimentally in
bacterial colonies -- including some that have only previously been
explained using far more complex models involving specific chemotactic
mechanisms.  Indeed, if motility decreases steeply enough with
density, then a spatially homogeneous bacterial population becomes
unstable to density fluctuations leading to the formation of bands
(1D) or droplets (2D). The length scale of the resulting pattern is
set by a balance between diffusion-drift fluxes and the logistic
relaxation of the population density towards its fixed-point value.
Starting instead from a small initial droplet of bacteria, we predict
the formation of concentric rings, each of which may eventually
further separate into droplets.

In several well studied systems, such characteristic patterns are
(with good reason) believed to be the direct result of chemotactic
behavior \cite{woodward,murray}. It is therefore remarkable that they
can also arise purely from the interplay of density-dependent
diffusivity and logistic growth, without explicit reference to the
dynamics (or even the presence) of a chemoattractant. \if{This
  suggests the possibility that a major role of the chemotactic
  signalling is to cause a reduction in motility at high density.}\fi
This suggests that similar patterns might arise in organisms having no
true chemotactic behavior at all. Such patterns could then be the
result of local chemical signalling without gradient detection
(quorum-sensing, not chemotaxis) or even purely physical interactions
(steric hindrance), either of which could in principle produce the
required dependence of motility on density. Last, a motility
decreasing with density is only one of the many mechanism that could
lead to ${\cal D}_{\rm e}'(\rho)<0$ and our analysis would apply
equally to all such cases.

The simplest version of our model allows identification of just two
dimensionless parameters that control the entire pattern-forming
process. In both homogeneous and centrosymmetric geometries, this
gives predictions for how the pattern type depends on interaction
strength which are broadly confirmed by experimental data.  This
suggests that some of the diverse patterns formed by colonies of
motile bacteria could have a relatively universal origin.

\begin{acknowledgments}
  We thank Otti Croze for discussions. We acknowledge funding from
  EPSRC EP/E030173. MEC holds a Royal Society Research Professorship.
  IP acknowledges the Spanish MICINN for financial support
  (FIS2008-04386).
\end{acknowledgments}

\appendix{}

\section{}
We show here how to derive the amplitude
equation~({8}). Let us start from the
dimensionless equation of motion~({6})
\begin{equation}
  \label{eqn:rescaledapp}
  \dot u= \nabla [ R e^{-2 \Phi u}(1-\Phi u)\nabla u]+u(1-u)-\nabla^4 u
\end{equation}
and recall the two conditions for patterning Eq.~({7}):
\begin{equation}
  \label{eqn:PDapp}
  \Phi>1;\qquad\qquad R \exp(-2\Phi)(\Phi-1)>2
\end{equation}
To analyze precisely the transition, we derive below the steady-state
limit of the amplitude equation in 1D. By inspection one sees that the
unperturbed steady-state of~\eqref{eqn:rescaledapp} is given by $u=1$. To
characterize the amplitude of the perturbation around $u=1$, we
introduce $u=1+w/\Phi$ so that $w$ evolves with
\begin{equation}
  \dot w = - \partial_x [R e^{-2 \Phi} (\Phi-1) (1+\frac{w}{\Phi -1}) e^{-2 w} \partial_x w] - w (1+\frac{w}{\Phi}) - \partial_x^4 w
\end{equation}
We are interested by the vicinity of the transition where
\begin{equation}
  R e^{-2\Phi} (\Phi-1)=2(1+\epsilon)
\end{equation}
for $\epsilon>0$ and small. The dynamics now reads
\begin{equation}
  \label{eqn:complete}
  \dot w = {\cal L} w - 2\epsilon \partial_x^2 w + g(w)
\end{equation}
where 
%\begin{equation}
  ${\cal L} = -(1+\partial_x^2)^2$ 
%\end{equation}
is the linear part of the evolution operator {\em at the transition},
$2\epsilon \partial_x^2 w$ gives an extra linear part due to the
perturbation ($\epsilon>0$) and $g(w)$ is the non-linear part:
\begin{equation}
  \label{eqn:NLterms}
  g(w)=-\frac{w^2}\Phi - \partial_x\Big[2(1+\epsilon) \big((1+\frac w{\Phi-1})e^{-2w}-1\big)\partial_x w\Big]
\end{equation}

\subsection{Amplitude equation}
As usual with the amplitude equation approach, we expand $w$ in power
series of the perturbation $\epsilon$ and study
Eq.~\eqref{eqn:complete} order by order. As shown below (Eqs
({21-23})), the correct expansion is
\begin{equation}
  \label{eqn:expofw}
  w=U_0 \epsilon^{1/2}+U_1 \epsilon+U_2 \epsilon^{3/2}+\dots
\end{equation}

\begin{figure*}
\includegraphics{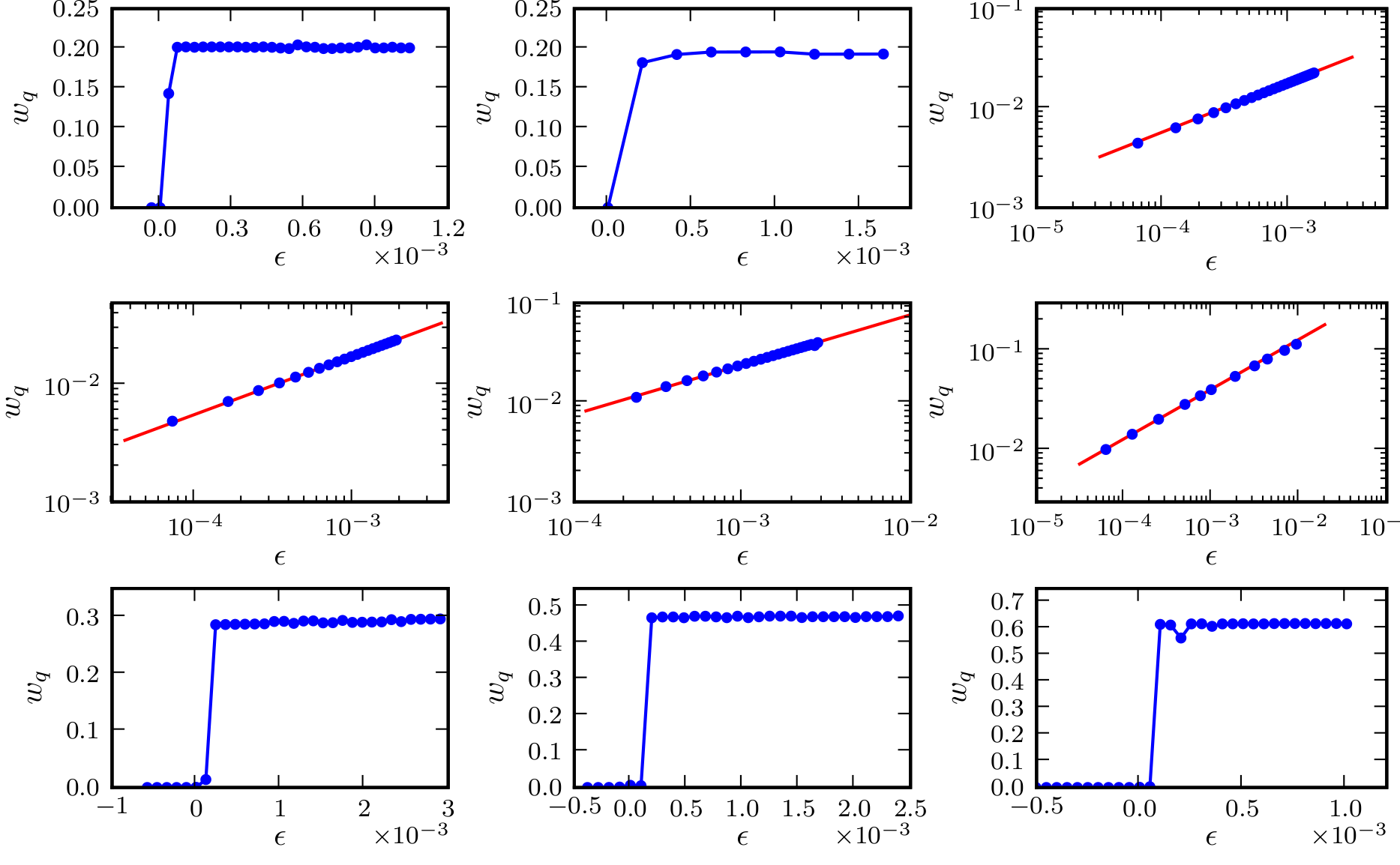}
\caption{We simulated equation~\eqref{eqn:rescaledapp} for systems of
size $L=400$ with periodic boundary conditions, using several values
of $\Phi$ (From left to right, top to bottom,
$\Phi=1.06;\,1.07;\,1.12;\,1.2;\,1.35;\,1.5;\,1.7;\,1.95;\,2.5$). The
steady-state $w(x)$ was then decomposed in Fourier series
$w(x)=a_0+\sum_{n}^{N/2} a_n \cos(2 \pi n x/L)+b_n \sin(2\pi n x/L)$,
where $N$ is the number of data points and the cut-off when
$n\to\infty$ is given by the Nyquist frequency. The blue points
correspond to the amplitude of the largest mode: $w_q={\rm
max}_n\sqrt{a_n^2+b_n^2}$. When the transition is continuous, we
compare these points with the results of the amplitude equation
$w_q=2\sqrt{\epsilon}|A|$, where $|A|$ is solution
of~\eqref{eqn:ampeq2} (red lines) and the agreement is excellent. For
$\Phi>1.58$ or $\Phi<1.08$, the transition is clearly discontinuous.}
\label{fig:amplitude}
\end{figure*}

Expanding~\eqref{eqn:NLterms} to the order $\epsilon^{3/2}$ and
substituting in~\eqref{eqn:complete} yields order by order:
\begin{eqnarray}
  -{\cal L} U_0&=&0\label{eqn:sqrteps}\\
  -{\cal L} U_1&=&-\frac{U_0^2}\Phi -\frac {3-2\Phi}{\Phi-1}\partial_x^2 U_0^2\label{eqn:eps}\\
  -{\cal L} U_2&=&-2 \partial_x^2 U_0 -\frac{2 U_0 U_1}\Phi-\frac{3-2\Phi}{\Phi-1} 2 \partial_x^2 U_0 U_1\label{eqn:eps3half}\\
    &&-\frac 4 3 \frac{\Phi-2}{\Phi-1} \partial_x^2 U_0^3\nonumber
\end{eqnarray}
Equation \eqref{eqn:sqrteps} can be easily solved and yields
\begin{equation}
  U_0=Ae^{ix}+A^* e^{-ix}
\end{equation}
The amplitude of the perturbation we are trying to derive is thus
 $2|A|$. Equation \eqref{eqn:eps} can also be solved directly:
\begin{equation}
  U_1=B e^{ix}+ B^* e^{-ix} + C + D e^{2ix} + D^* e^{-2ix}
\end{equation}
where $B$ can de determined from higher order equations (but does not
interest us here), and $C$ and $D$ are given by
\begin{equation}
  C=-\frac {2 |A|^2}\Phi ;\qquad D=\frac{A^2}9(4 \frac{3-2\Phi}{\Phi-1}-\frac 1 \Phi)
\end{equation}
as can be checked by direct substitution in Eq.~\eqref{eqn:eps}. Equation
\eqref{eqn:eps3half} does not always have a solution. Indeed, the
application of ${\cal L}$ to any function $U_2$ cannot yield a
multiple of $e^{ix}$, (since ${\cal L} e^{ix}=0$ and ${\cal L}$ is
linear). The r.h.s. however does contain a multiple of $e^{ix}$ whose
prefactor must thus vanish. This gives a condition for the expansion
to provide a proper steady-state solution of the problem. Let us
summarize the contributions of the different terms to the prefactor of
$e^{ix}$ in the r.h.s of Eq.~\eqref{eqn:eps3half}
\begin{eqnarray}
  -2 \partial_x^2 U_0 & {\rm yields}& 2 A\label{eqn:contriblineareps}\\
  (\frac{3-2\Phi}{\Phi-1} 2 \partial_x-\frac{2 }\Phi) U_0 U_1&{\rm yields}&2 A |A|^2 \Big(\frac{3-2\Phi}{\Phi-1}-\frac 1 \Phi\Big)\\
  &&\times\Big(\frac 1 9 (4\frac{3-2\Phi}{\Phi-1}-\frac 1 \Phi)-\frac 2 \Phi\Big)\nonumber\\
  -\frac 4 3 \frac{\Phi-2}{\Phi-1} \partial_x^2 U_0^3&{\rm yields}& 4 \frac{\Phi-2}{\Phi-1} A|A|^2
\end{eqnarray}
The sum of these terms vanishes only if
\begin{equation}
  \label{eqn:ampeq1}
  A\Big( 9\Phi^2 (\Phi-1)^2+2 |A|^2 (34 \Phi^4-56 \Phi^3-24\Phi^2+31\Phi+19)\Big)=0
\end{equation}
and thus either
\begin{equation}
  \label{eqn:ampeq2}
  A=0;\qquad\mbox{or}\qquad |A|^2=\frac{9 \Phi^2 (1-\Phi)^2}{2(34 \Phi^4-56 \Phi^3-24\Phi^2+31\Phi+19)}
\end{equation}
Finally, the first order in the amplitude equation yields
\begin{equation}
  \label{eqn:ampeq3}
  w(x)=2|A|\sqrt{\epsilon} \cos(x-x_0)
\end{equation}
where $x_0$ is a constant. Note that by construction $|A|^2>0$ and a
non-zero solution only exists for $\Phi\in[1.08439, 1.59237]$. For
these values of $\Phi$, Eq.~\eqref{eqn:ampeq2} and~\eqref{eqn:ampeq3}
work very well, as can be checked in figure
\ref{fig:amplitude}. Outside this range the transition becomes
subcritical and the standard approach does not work
anymore. Alternative treatments have been proposed but are not as
reliable (see ref~[{20}] for more details). Interestingly, we see
that the order of the transition and the amplitude of the perturbation
depend on how non-linear terms in $g(w)$ balance the linear growth
term $-2 \epsilon \partial_x^2 w$ in \eqref{eqn:complete}. Since the
former depends on the non-linear relation $v(\rho)$, we do not expect
equation~\eqref{eqn:ampeq2} to be generic, as opposed to the stability
analysis which can be expressed solely in terms of ${\cal D}_{\rm
  e}(\rho_0)$ and its derivative.

\subsection{What is the correct expansion?}
\label{sec:order}

In~\eqref{eqn:expofw}, we expanded $w$ in power series of
$\sqrt{\epsilon}$, thus assuming that the amplitude is an analytic
function of $\sqrt{\epsilon}$. One could look for a more general
expansion:
\begin{equation}
  w=U_0 \epsilon^\alpha+U_1 \epsilon^{2\alpha}+U_2 \epsilon^{3\alpha}
\end{equation}
In this case, the expansion of equation \eqref{eqn:complete} yields
two power series: $\sum R_k \epsilon^{\alpha k}$ and $\sum W_k
\epsilon^{\alpha k+1}$. For the two series to give terms that can
balance each-other, one needs
%\begin{equation}
  $\alpha+1=k\alpha$ for $k\geq 2$ 
%\end{equation}
and thus
\begin{equation}
  \alpha=\frac{1}{k-1}
\end{equation}
The candidates for $\alpha$ are thus $1;\,1/2;\,1/3;\,\dots$. Note
that $\alpha \leq 1$ implies $2\alpha+1\geq 3 \alpha$. We can
therefore stop the expansion at $3\alpha$ and $2 \alpha+1$ to get the
first three terms in the expansion of equation \eqref{eqn:complete}

Let us first try $\alpha=1$. The order by order the expansion yields
\begin{eqnarray}
  L^2 U_0&=&0,\quad {\cal O}(\epsilon)\label{eqn:eq11}\\
  L^2 U_1&=&-\frac{U_0^2}\Phi -\frac{3-2\Phi}{\Phi-1} \partial_x^2 U_0^2 -2 \partial_x^2 U_0,\quad {\cal O}(2\epsilon)\label{eqn:eq12}
\end{eqnarray}
Equation \eqref{eqn:eq11} yields $U_0=Ae^{ikx}+A^* e^{-ikx}$ but
equation \eqref{eqn:eq12} cannot be solved since there is a non-zero
multiple of $e^{ikx}$ on the r.h.s. ($-2 \partial_x^2 U_0$) which
cannot result from the application of $L^2$ to any function. Thus
$\alpha=1$ is not an option.

For $\alpha\leq 1/3$, then $\alpha+1>1\geq 3 \alpha$. There is thus no
contribution of $-2 \epsilon \partial_x^2 w$ to the first three orders
in the expansion of~\eqref{eqn:complete}. In particular, the two first
order are still given by~\eqref{eqn:sqrteps} and~\eqref{eqn:eps},
whereas the third order is given by~\eqref{eqn:eps3half} without the
term linear in $U_0$. This means that the contribution
\eqref{eqn:contriblineareps} is not present and the prefactor of
$e^{ikx}$ in the r.h.s. of \eqref{eqn:eps3half} only contains
multiples of $|A|^2 A$. The resolvability condition~\eqref{eqn:ampeq1}
is thus of the form $A|A|^2 \,f(\Phi)=0$ which implies $|A|=0$. The
only expansion which yields a result is thus for $\alpha=1/2$.

\end{document}